\documentclass[aps,floatfix,twocolumn,prb,superscriptaddress,reprint,amsmath,amssymb]{revtex4-2}

\usepackage{graphicx}
\usepackage{dcolumn}
\usepackage{bm}

\usepackage[utf8]{inputenc}
\usepackage[T1]{fontenc}
\usepackage{mathptmx}
\usepackage{color}
\usepackage{etoolbox}
\usepackage{braket}
\usepackage{amsmath}
\usepackage{dsfont}
\usepackage{mathtools}
\usepackage{comment}
\usepackage{linop}

\newcommand{\abs}[1]{\ensuremath{\left\vert#1\right\vert}}

\renewcommand{\arraystretch}{1.5}
\newcommand{\nh}[1]{#1}

\usepackage{xcolor}
\definecolor{chalmersvastkust}{RGB}{0,48,80}
\usepackage{hyperref}
\hypersetup{
  colorlinks,
  citecolor=chalmersvastkust,
  linkcolor=chalmersvastkust,
  urlcolor=chalmersvastkust}

\begin{document}

\title[]{Collective Buckling in Metal-Organic Framework Materials}

\author{Nico Hahn}
\affiliation{Department of Physics, Chalmers University of Technology, Gothenburg, Sweden}
\email{nico.hahn@chalmers.se}

\author{Lars Öhrström}
\affiliation{Department of Chemistry and Chemical Engineering, Chalmers University of Technology, Gothenburg, Sweden}

\author{R. Matthias Geilhufe}
\affiliation{Department of Physics, Chalmers University of Technology, Gothenburg, Sweden}

\date{\today}

\begin{abstract}
We develop a framework to describe collective buckling in metal-organic frameworks (MOFs). Starting from the microscopic structure of a single organic linker, we define a buckling coordinate governed by an effective double-well potential. \nh{Coupling between linkers is introduced within a dipole–dipole approximation, resulting in an effective lattice Hamiltonian.} We analyze the transition between ordered and disordered phases within a mean-field approximation and \nh{estimate} the critical temperature. As an \nh{illustrative} example for our theory, we discuss the collective buckling instability for the prototypical cubic framework MOF-5 under different values of uniaxial strain. Our approach \nh{provides} a quantitative description of collective buckling in framework materials. 
\end{abstract}

\maketitle

\section{Introduction}
\label{secI}

Buckling is a ubiquitous phenomenon in applied mechanics, in which structural elements adopt deformed configurations under sufficient axial load \cite{BazantCedolin2010, Guerra2023}. Such a transition involves the spontaneous breaking of a symmetry and leads to the emergence of new stable states. Similar mechanisms are also possible at the molecular scale, for example in the organic linkers of metal-organic frameworks (MOFs) \nh{\cite{Schneemann2014, Coudert2015, Ying2021, Pallach2021}}.

MOFs are relatively new materials consisting of metallic coordination centers connected by organic linkers, often forming porous crystalline structures with large internal surface area \cite{IUPAC}. Their flexible design has led to more than 90.000 discovered structures with diverse functionalities. Technological applications include gas storage and separation, catalysis and chemical sensing, among others \cite{LarsFrancoiseBook, Furukawa2013}.

Beyond that MOFs also provide a promising platform for quantum materials \cite{HuangGeilhufe2024}. To date, phenomena such as topological electronic states \cite{Jiang2021}, strong correlations \cite{Kumar2021, Field2022, Lowe2024} and superconductivity \cite{Haung2018, Takenaka2021, Ohlrich2025} have been explored in both theoretical and experimental works. While conventional quantum materials exhibit macroscopic order composed of microscopic degrees of freedom such as spin, charge or orbital, the unconventional degrees of freedom in MOFs, in combination with their flexible design, can give rise to distinct types of ordering phenomena. In this work, we demonstrate one example: a collective buckling transition.

For a single linker, buckling typically leads to the emergence of multiple stable configurations related by a symmetry operation. In framework materials, however, the buckling state of each linker is influenced by its neighbours. \nh{In principle,} such interactions \nh{can} give rise to collective phases reminiscent of magnetic ordering \cite{Geilhufe2021}. In the ferrobuckling phase, linkers tend to buckle in the same direction, while in the antiferrobuckling phase they tend to buckle in opposite directions. \nh{At low temperatures, quantum fluctuations can suppress the transition into the classically ordered state, giving rise to a parabuckling phase~\cite{Geilhufe2021}, where tunneling between symmetry-related configurations dominates over interactions and the linkers remain in a superposition of buckling states without selecting a definite orientation. This is analogous to the quantum paraelectric state observed in perovskites such as SrTiO$_3$~\cite{Rowley2014, Esswein2022}.}

\begin{figure}[t]
    \centering
    \includegraphics[width=0.48\textwidth]{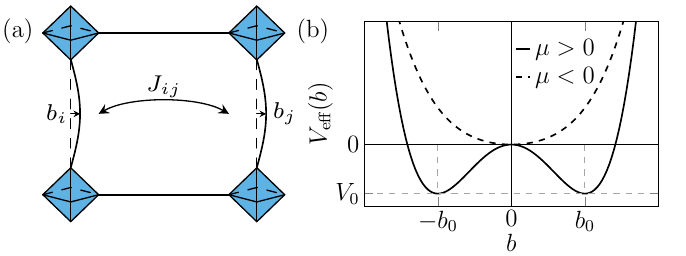}
    \caption{Molecular buckling and \nh{dipolar coupling} in MOFs. (a) Schematic illustration of a MOF composed of \nh{metallic coordination centers}, shown as tetrahedra, and molecular linkers shown as solid lines. The amplitude of the buckling is described by the scalar quantity $b$, the mode amplitude. The interaction \nh{between} molecules, denoted by $J_{ij}$, is modeled as a dipole-dipole \nh{coupling}. (b) Effective potential of molecular buckling. Parabolic behaviour (dashed line) describes straight molecules without buckling. The double-well potential (solid line) \nh{favors} molecular buckling.}
    \label{fig:schematic}
\end{figure}

\begin{figure}
    \centering
    \includegraphics[width=0.45\textwidth]{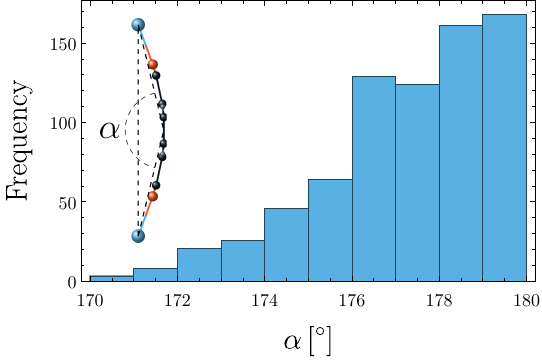}
    \caption{Distribution of linker angles for the 1,4-benzenedicarboxylate (bdc) linker of the metal-organic framework MOF-5, obtained from the Cambridge Structural Database (CSD) \cite{CSD}. \nh{The inset defines the linker angle $\alpha$.} An angle of 180$^{\circ}$ corresponds to a straight unbuckled configuration, while smaller angles indicate increasing buckling. The plot stops at the angle 179.5$^{\circ}$ as the large majority of structures have a symmetry restricted angle of exactly 180$^{\circ}$.}
    \label{fig:BucklingHistogram}
\end{figure}

The molecular buckling in MOFs is schematically illustrated in Fig.~\ref{fig:schematic}. The \nh{metallic coordination centers, or secondary building units (SBUs),} are arranged in a lattice and connected to their nearest neighbours by organic linkers. The framework responds to external stress through a shift of the \nh{metallic centers}, which in turn induces a buckling of the organic linkers. In Fig.~\ref{fig:BucklingHistogram} we show the distribution of linker angles in the 1,4-benzenedicarboxylate (bdc) linker of MOF-5 \cite{Li1999} obtained from 750 datasets of the Cambridge Structural Database (CSD) \cite{CSD}.

We show that the molecular buckling in MOFs can be described classically by the lattice Hamiltonian
\begin{equation}    \label{1Ham}
    H = - \frac{1}{2} \sum_{i,j} J_{ij} b_i b_j + \sum_i V_{\text{eff}}(b_i)
\end{equation}
with an effective double-well potential 
\begin{equation}
    V_{\text{eff}}(b) = -\mu\, b^2 + \frac{\nu}{2} b^4,
\end{equation}
see Fig.~\ref{fig:schematic}, and a real-valued buckling coordinate $b \in \mathbb{R}$. \nh{The parameters $\mu$ and $\nu$ depend on the applied strain.} For $\mu > 0$ there are two degenerate minima corresponding to a buckling to the left or the right. For $\mu \leq 0$ there is no stable buckling configuration. \nh{The interaction between linkers $i$ and $j$ is introduced at the level of a minimal microscopically motivated dipole-dipole coupling and is linear in both buckling coordinates.}

\nh{This description constitutes a coarse-grained model that focuses on a specific low-energy deformation mode of the framework. Alternative collective distortions of MOF-5, in particular displacements of the metallic coordination centers associated with angular distortions of the cubic network, have been identified experimentally \cite{Pallach2021} and are reflected in an anomalously soft transverse acoustic phonon branch \cite{Rimmer2014}. The molecular buckling coordinate adopted here provides a complementary coarse-grained description that emphasizes the symmetry-breaking character of linker deformations and their collective ordering.}

\nh{A collective buckling transition suggests possible applications for framework materials by promoting cooperative mechanically driven structural responses. It enables a route towards mechanically controlled porosity and, consequently, mechanically driven changes in adsorption and diffusion properties \cite{Coudert2015, Ying2021}. Moreover, the coupling of electronic degrees of freedom to buckling deformations suggests possible routes towards mechanically tunable electronic or electromechanical phenomena \cite{Savelev2006, Geilhufe2021}.}

The paper is organized as follows. In Sec.~\ref{secII}, we present a rigorous derivation of the buckling coordinate of a single linker. In Sec.~\ref{secIII}, we derive the \nh{coupling} $J_{ij}$ microscopically by approximating each linker as an electric dipole. These two ingredients lead to the lattice model \eqref{1Ham}. In Sec.~\ref{secIV}, we consider the collective buckling phases of this model within a mean-field approximation. The technical details of this section are relegated to the appendix. \nh{In Sec.~\ref{secV}, we illustrate our approach using the metal-organic framework MOF-5 as a well-characterized and widely studied example, for which we obtain the model parameters via density functional theory (DFT). This application is intended to demonstrate the resulting collective buckling physics rather than to provide a complete description of MOF-5.} In Sec.~\ref{secVI}, we discuss the classical to quantum cross-over at low temperatures and recover the transverse field Ising model, which has been considered in the context of buckling in Ref.~\onlinecite{Geilhufe2021}. Finally, we conclude in Sec.~\ref{secVII}.

\section{The Buckling Coordinate}
\label{secII}

In this section, we derive the collective buckling coordinate of a molecular linker and map the buckling dynamics onto an effective single-particle problem characterized by an effective potential.

Let $\mathbf{x}_i \in \mathbb{R}^3$ be the coordinates of the $i$th ($i = 1, \ldots, N$) atom and $V(\mathbf{x}_1, \ldots, \mathbf{x}_N)$ the potential of the unbuckled molecule. The equilibrium configuration of the molecule $(\mathbf{x}_1^{(0)}, \ldots, \mathbf{x}_N^{(0)})$ is a local minimum of the potential. Without loss of generality, we set the end points to $\mathbf{x}_1^{(0)} = 0$ and $\mathbf{x}_N^{(0)} = L\, \hat{e}_1$, where $L$ is the length of the molecule and $\hat{e}_1$ an arbitrary unit vector indicating the chain direction. 

Buckling is introduced by constraining the endpoints to
\begin{equation}
    \mathbf{x}_1 = \mathbf{x}_1^{(0)} = 0, \qquad  \mathbf{x}_N = L (1-\kappa) \hat{e}_1,
\end{equation}
where the parameter $\kappa$ with $0 \leq \kappa \ll 1$ quantifies the imposed \nh{strain}. In the case of multiple atoms at the ends of the chain, we assume that they are rigid and define $\mathbf{x}_1$ and $\mathbf{x}_N$ as their geometric centers. \nh{In this description, the positions of the metallic centers are treated as fixed and define geometric boundary conditions for the linkers. Displacements of the metallic centers are not included as explicit degrees of freedom.} This defines a constrained potential
\begin{align}
    V_\kappa (\mathbf{x}) &= V_\kappa (\mathbf{x}_2, \ldots, \mathbf{x}_{N-1}) \notag \\
    &= V (\mathbf{x}_1 = 0, \mathbf{x}_2, \ldots, \mathbf{x}_{N-1}, \mathbf{x}_N = L (1-\kappa) \hat{e}_1),
\end{align}
which depends only on the inner coordinates $\mathbf{x} = (\mathbf{x}_2, \ldots, \mathbf{x}_{N-1}) \in \mathbb{R}^{3(N-2)}$. \nh{The equilibrium in the unbuckled configuration with zero strain ($\kappa = 0$) serves as a reference and is denoted as $\mathbf{x}^{(0)} = (\mathbf{x}_2^{(0)}, \ldots, \mathbf{x}_{N-1}^{(0)})$.}

\begin{figure}
    \centering
    \includegraphics[width=0.4\textwidth]{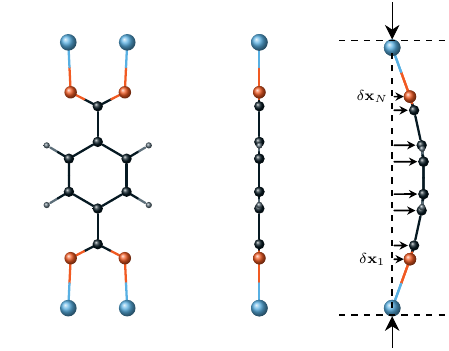}
    \caption{Illustration of the atomic displacement for the bdc linker of MOF-5. The relaxed configuration is shown on the left and in the center. The buckled configuration is shown on the right with atomic displacements $\mathbf{\delta x}_j$.}
    \label{fig:bucklingcoordinate}
\end{figure}

\nh{For sufficiently small strain, the unbuckled configuration remains stable. A buckling instability occurs only beyond a critical strain $\kappa_c$. For small $\kappa - \kappa_c > 0$, the equilibrium position will shift by $\boldsymbol{\delta} \mathbf{x}$.} Fig.~\ref{fig:bucklingcoordinate} illustrates the atomic displacement due to buckling. We find this displacement vector by expanding the potential to first order around $\mathbf{x}^{(0)}$ and equating to zero
\begin{equation}    \label{2TaylorExp}
    \nabla V_\kappa (\mathbf{x}^{(0)} + \boldsymbol{\delta} \mathbf{x}) = \nabla V_\kappa (\mathbf{x}^{(0)}) + \mathcal{H} \boldsymbol{\delta} \mathbf{x} + \mathcal{O}\left( \boldsymbol{\delta} \mathbf{x}^2 \right) \stackrel{!}{=} 0,
\end{equation}
where
\begin{equation}
    \mathcal{H}_{i\alpha, j\beta} = \frac{\partial^2 V_\kappa (\mathbf{x})}{\partial x_{i\alpha} \partial x_{j\beta}} \Bigg|_{\mathbf{x} = \mathbf{x}^{(0)}}
\end{equation}
is the Hessian of the potential at the equilibrium. For the sake of clarity, we do not include the dependency on $V_\kappa(\mathbf{x})$ into the notation of the Hessian. Furthermore, we follow the convention that Latin indices ($i,j, \ldots$) label atoms and Greek indices ($\alpha,\beta, \ldots$) denote Cartesian coordinates. The displacement vector is thus given by
\begin{equation}
    \boldsymbol{\delta} \mathbf{x} = - \mathcal{H}^{-1} \nabla V_\kappa (\mathbf{x}^{(0)}).
\end{equation}
Before defining the buckling coordinate, it is convenient to switch to mass-weighted coordinates
\begin{equation}    \label{2MassCoordinates}
    \mathbf{q} = M^{1/2} \mathbf{x},
\end{equation}
where $M = \text{diag}(m_2, \ldots, m_{N-1}) \otimes \mathds{1}_3$ is the diagonal mass matrix. This transformation is also used in the discussion of vibrational modes and phonons. The corresponding mass-weighted Hessian
\begin{equation}    \label{2MassHessian}
    \mathcal{D} = M^{-1/2} \mathcal{H} M^{-1/2}
\end{equation}
retains its real symmetric form. We denote the eigenvalues as $\lambda^{(i)}$ and the normalized eigenvectors as $\mathbf{v}^{(i)}$ and order them as $\lambda^{(1)} \leq \lambda^{(2)} \leq \ldots \leq \lambda^{(3(N-2))}$. The eigenvector $\mathbf{v}^{(1)}$ belonging to the smallest eigenvalue defines the soft buckling mode, i.e. the direction of the instability. Here we assume that the eigenvalues are non-degenerate. A comment on this point follows at the end of this section.

The scalar buckling coordinate is defined as the projection of the displacement $\boldsymbol{\delta} \mathbf{q}$ (now in mass-weighted coordinates) onto the normalized soft eigenmode
\begin{equation}
    b = \mathbf{v}^{(1)} \cdot \boldsymbol{\delta} \mathbf{q} = - \frac{1}{\lambda^{(1)}} \mathbf{v}^{(1)} \cdot M^{-1/2} \nabla V_\kappa (\mathbf{x}^{(0)}),
\end{equation}
which is measured in units of $[b] = \sqrt{\text{u}}\, \text{\AA}$. Conversely, the displacement vector can be written in a single-mode approximation as
\begin{equation}    \label{2SoftModeApprox}
    \boldsymbol{\delta} \mathbf{q} \approx b\, \mathbf{v}^{(1)}.
\end{equation}
The single-mode approximation is good when the gap between the smallest and the second smallest eigenvalue is large, $\lambda^{(2)} \gg \lambda^{(1)}$. We define the effective potential as
\begin{equation}
    V_{\text{eff}}(b) = V_\kappa \left( \mathbf{x}^{(0)} + b\, M^{-1/2} \mathbf{v}^{(1)} \right).
\end{equation}
The relevant physics is captured by the low-energy expansion, the double-well potential
\begin{equation}    \label{2EffPot}
    V_{\text{eff}}(b) = -\mu\, b^2 + \frac{\nu}{2} b^4 + \mathcal{O} \left( b^6 \right),
\end{equation}
where we ignore the constant energy shift. \nh{This low-energy expansion is intended to describe the regime of small to moderate buckling coordinates. In this regime, the atomic displacements remain small compared to the molecular length, providing an a-posteriori consistency check for both \eqref{2TaylorExp} and \eqref{2EffPot}. Large buckling coordinates lie outside the scope of this model.

The parameter $\mu$ is related to the curvature of the potential along the soft mode and indicates the presence of a buckling instability, while $\nu$ stabilizes the potential at a finite buckling coordinate. Applied strain modifies the potential and therefore enters the model through a dependence of these parameters.} Henceforth, we will not explicitly state terms of $\mathcal{O}(b^6)$, which become important only in the case $\nu \leq 0$.

The transformation to mass-weighted coordinates \eqref{2MassCoordinates} guarantees that the effective mass of the buckling coordinate is unity. This can be checked quickly by considering the kinetic energy of the system in the soft mode approximation \eqref{2SoftModeApprox}
\begin{equation}
    T = \frac{1}{2} \dot{\mathbf{x}}^T M \dot{\mathbf{x}} = \frac{1}{2} \dot{\mathbf{q}}^2 \approx \frac{1}{2} \dot{b}^2.
\end{equation}

Finally, we discuss the role of symmetry for the soft mode. Let $G$ denote the point group that leaves the equilibrium configuration $\mathbf{x}^{(0)}$ invariant. The group elements $g \in G$ act on the displacement $\boldsymbol{\delta} \mathbf{x}$ via an orthogonal matrix $R(g)$
\begin{equation}
    \boldsymbol{\delta} \mathbf{x} \to R(g) \boldsymbol{\delta} \mathbf{x}.
\end{equation}
Since the potential at the equilibrium is invariant, the Hessian commutes with all $R(g)$
\begin{equation}
    \mathcal{H} = R(g) \mathcal{H} R^T(g).
\end{equation}
Therefore the Hessian becomes block-diagonal in the basis where $R(g)$ is reduced to its irreducible components. The same holds true for the mass-weighted Hessian \eqref{2MassHessian} as long as the mass distribution respects the symmetry, i.e. $M$ commutes with all $R(g)$. As a consequence of Schur's lemma, the symmetry enforces degeneracies that coincide with the dimensions of the real irreducible representation of $G$.

In particular, $\mathbb{Z}_2$ has only one-dimensional irreducible representations and thus the eigenvalues of the Hessian are non-degenerate. In contrast, higher symmetry groups such as $C_n \cong \mathbb{Z}_n$ ($n \geq 3$) or $\rm SO(2)$ have also two-dimensional real irreducible representations, which enforces twofold degeneracies of the soft mode. This latter case requires two buckling coordinates, defined as the projections of $\boldsymbol{\delta} \mathbf{x}$ onto the two eigenvectors belonging to the smallest eigenvalue. In this situation, our formalism applies in a similar manner, but details are beyond the scope of this work.

\section{Dipolar Approximation of Linker-Linker Interaction}
\label{secIII}

Each linker carries a charge distribution $\rho(\mathbf{r}; \mathbf{x})$ with zero net charge. The distribution has an ionic contribution depending directly on the atomic coordinates $\mathbf{x}$, and thus on the buckling coordinate, and a continuous contribution of the electrons relaxing around these positions. The dipole moment of a linker in the configuration $\mathbf{x}$ is
\begin{equation}    \label{3DipoleMomentDef}
    \mathbf{p}(\mathbf{x}) = \int d^3r\, \rho(\mathbf{r}; \mathbf{x}) \mathbf{r}.
\end{equation}
The equilibrium position changes in the presence of buckling and we expand
\begin{equation}
    \mathbf{p} \left( \mathbf{x}^{(0)} + \boldsymbol{\delta} \mathbf{x} \right) = \mathbf{p}\left( \mathbf{x}^{(0)} \right) + Z\, \boldsymbol{\delta} \mathbf{x} + \mathcal{O}\left( \boldsymbol{\delta} \mathbf{x}^2 \right),
\end{equation}
where we introduced the effective charge tensor
\begin{equation}    \label{3EffCharge}
    Z_{\alpha,i \beta} = \frac{\partial p_\alpha(\mathbf{x})}{\partial x_{i \beta}}\Bigg|_{\mathbf{x} = \mathbf{x}^{(0)}}.
\end{equation}
This quantity is the molecular analogue of the Born effective charge defined for the polarization per unit cell~\cite{Ghosez1998,Gonze1997}.

The atomic displacement is dominated by the soft buckling mode \eqref{2SoftModeApprox} and we obtain
\begin{equation}    \label{3DipoleApprox}
    \mathbf{p} = \mathbf{p}^{(0)} + b\, \mathbf{d},    \qquad  d_\alpha =  \sum_{i, \beta} \frac{Z_{\alpha,i \beta}}{\sqrt{M_i}} v^{(1)}_{i \beta},
\end{equation}
where we suppress the argument of $\mathbf{p}$ and use the abbreviation $\mathbf{p}^{(0)} = \mathbf{p}(\mathbf{x}^{(0)})$. We define a mode dipole vector $\mathbf{d}$ and take the mass-weighted coordinates into account. The dipole vector is measured in units of $[d] = e/\sqrt{\text{u}}$, which combines to units of a true dipole moment when multiplied with the buckling coordinate.

We label different linkers by $i$ and $j$ and consider their dipole-dipole interaction
\begin{equation}
    E_{ij} = \frac{1}{4\pi \epsilon} \frac{\mathbf{p}_i \cdot \mathbf{p}_j - 3 \left( \mathbf{p}_i \cdot \hat{e}_{ij} \right) \left( \mathbf{p}_j \cdot \hat{e}_{ij} \right)}{r_{ij}^3},
\end{equation}
where $\epsilon$ is the permittivity of the material, \nh{$r_{ij}$ is the distance between linkers $i$ and $j$ and $\hat{e}_{ij}$ the normalized center-to-center vector connecting the linkers.}

Inserting \eqref{3DipoleApprox} we obtain
\begin{equation}    \label{3DipoleInteraction}
    E_{ij} =  -J_{ij}^{(0)} - J_{ij}^{(1)} b_i - J_{ji}^{(1)} b_j - J^{(2)}_{ij} b_i b_j
\end{equation}
with
\begin{equation}
\begin{split}
    J^{(0)}_{ij} &= \frac{3 (\mathbf{p}_i^{(0)} \cdot \hat{e}_{ij} ) ( \mathbf{p}_j^{(0)} \cdot \hat{e}_{ij} ) - \mathbf{p}_i^{(0)} \cdot \mathbf{p}_j^{(0)}}{4\pi \epsilon\, r_{ij}^3},
    \\
    J^{(1)}_{ij} &= \frac{3 ( \mathbf{d}_i \cdot \hat{e}_{ij} ) ( \mathbf{p}_j^{(0)} \cdot \hat{e}_{ij} ) - \mathbf{d}_i\cdot \mathbf{p}_j^{(0)}}{4\pi \epsilon\, r_{ij}^3},
    \\
    J^{(2)}_{ij} &= \frac{3 ( \mathbf{d}_i \cdot \hat{e}_{ij} ) ( \mathbf{d}_j \cdot \hat{e}_{ij} ) - \mathbf{d}_i \cdot \mathbf{d}_j}{4\pi \epsilon\, r_{ij}^3},
\end{split}
\end{equation}
where we included a sign so that \nh{the interaction parameters are positive} for an attractive interaction. The interaction is symmetric $J^{(k)}_{ij} = J^{(k)}_{ji}$ and we take $J^{(k)}_{ii} = 0$ for $k \in \{0, 1, 2\}$. The parameters are expressed in units of $[J^{(0)}_{ij}] = \text{eV}$, $[J^{(1)}_{ij}] = \text{eV}/(\sqrt{\text{u}}\, \text{\AA})$ and $[J^{(2)}_{ij}] = \text{eV}/(\text{u}\, \text{\AA}^2)$. Eq. \eqref{3DipoleInteraction} simplifies tremendously when $\mathbf{p}^{(0)} = 0$ for all linkers
\begin{equation}    \label{3DipoleInteraction2}
    \nh{E_{ij} = - J^{(2)}_{ij} b_i b_j.}
\end{equation}
This is the case when the linkers have spatial inversion symmetry, since then we have $\rho(\mathbf{r}; \mathbf{x}) = -\rho(-\mathbf{r}; -\mathbf{x})$, so that a vanishing dipole moment follows immediately from \eqref{3DipoleMomentDef}. 

\nh{The resulting bilinear term provides a minimal microscopically motivated description of linker-linker coupling arising from electrostatic interactions. Additional coupling mechanisms, such as elastic constraints mediated by the metallic centers or higher-order multipole contributions, can be incorporated within extended effective models.}

\section{Macroscopic Phase Transition}
\label{secIV}

Taking the \nh{effective dipolar coupling between linkers} and the local effective potential into account, we obtain the lattice Hamiltonian \eqref{1Ham}
\begin{equation}    \label{4Ham}
    H = -\frac{1}{2} \sum_{i,j} J_{ij} b_i b_j + \sum_i V_{\text{eff}}(b_i),
\end{equation}
where we assume the situation of linkers with spatial inversion symmetry and have included a factor $1/2$ to avoid double counting. \nh{For convenience of notation, we drop the superscript of the coupling constant \eqref{3DipoleInteraction2}.}

In analogy to magnetization in spin models, we define a global buckling order parameter by
\begin{equation}
    m = \frac{1}{N} \sum_i \langle b_i \rangle = \langle b \rangle,
\end{equation}
which is normalized by the system size $N$. Since we consider bulk properties, there is translation invariance and the order parameter reduces to the thermal expectation value $\langle b \rangle$ of any linker, indicated by the absence of a site index.

We apply a mean-field approximation
\begin{equation}
    b_i = \langle b_i \rangle + \delta b_i, 
\end{equation}
i.e. each buckling coordinate is written as its average plus thermal fluctuations $\delta b_i = b_i - \langle b_i \rangle$. In the mean-field Hamiltonian, all but the leading order of $\delta b_i$ are neglected. It then becomes an effective single-particle Hamiltonian
\begin{equation}    \label{4HamMF}
    H_{\text{MF}} = \sum_i h(b_i) + \frac{N \Tilde{J}}{2} m^2
\end{equation}
with a local Hamiltonian
\begin{equation}
    h(b_i) = -\Tilde{J}\, m\, b_i + V_{\text{eff}}(b_i)
\end{equation}
and
\begin{equation}    \label{4MFCoupling}
    \Tilde{J} = \sum_{j} J_{0j}.
\end{equation}
The order parameter is determined by the self-consistency equation of the mean-field theory
\begin{equation}
    m = \frac{\displaystyle \int d[b] b_i\, e^{-\beta H_{\text{MF}}}}{\displaystyle \int d[b] e^{-\beta H_{\text{MF}}}} = \frac{\displaystyle \int db\, b\, e^{-\beta h(b)}}{\displaystyle \int db\, e^{-\beta h(b)}}
\end{equation}
where we use the notation $d[b] = db_1 \cdots db_N$. Since we are dealing with an effective single-particle theory, the integral factorizes and we are left with a single integral over a buckling coordinate.

In Appendix~\ref{A1}, we perform a cumulant expansion of the free energy and obtain the critical temperature as the solution of the fixed-point equation
\begin{equation}
    k_\mathrm{B} T_C = \Tilde{J} \left\langle b^2 \right\rangle_0,
\end{equation}
where
\begin{equation}    \label{4SinglePartPartFunc}
    \left\langle f(b) \right\rangle_0 = \frac{1}{Z_0} \int db\, f(b) e^{-\beta V_{\text{eff}}(b)},
    \qquad
    Z_0 = \int db\, e^{-\beta V_{\text{eff}}(b)}
\end{equation}
denotes the thermal expectation value of a single linker in the absence of interactions. The moments, such as $\langle b^2 \rangle_0$, can be evaluated analytically in terms of parabolic cylinder functions, providing a closed-form description of the mean-field theory.

\section{Application to MOF-5}
\label{secV}

\begin{figure}[t]
    \centering
    \includegraphics[width=0.49\textwidth]{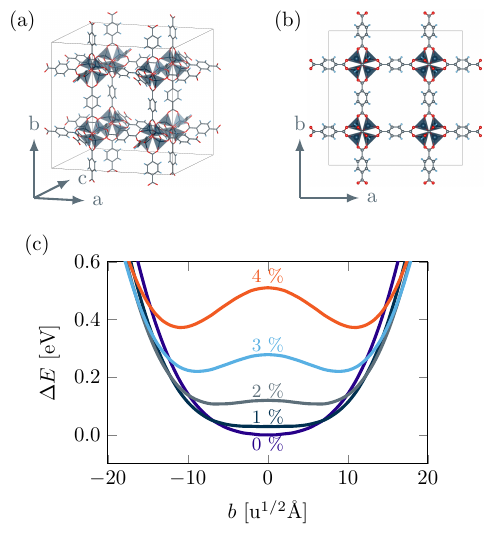}
    \caption{Buckling of bdc molecules in MOF-5. \nh{Crystallographic directions are indicated by the cubic lattices axes $a$, $b$ and $c$.} (a) The crystal
    structure of MOF-5. Zn$_4$O \nh{centers} are coordinated by bdc molecules, forming a periodic three-dimensional cubic network. (b) Top view of a (100)-plane
    (c) Total energy versus the buckling of bdc
    molecules for various strain values. Under strain, the total energy follows the form of a double-well potential.}
    \label{fig:potential}
\end{figure}

Having established the mean-field framework, we now \nh{consider} a specific material \nh{as an illustrative example}. MOF-5 is a paradigmatic metal-organic framework with the chemical sum formula [Zn$_4$O(bdc)$_3$], where bdc stands for 1,4-benzenedicarboxylate \cite{Li1999}. It crystallizes in a cubic lattice, with a nearest-neighbour distance of 12.9~\AA~between two zinc oxide \nh{centers} (see Fig.~\ref{fig:potential}(a) and (b)). Due to an alternating torsion of \nh{the} bdc molecules, the lattice constant is twice as large. MOF-5 is a soft material with a calculated bulk modulus of 15.37~GPa~\cite{Yang2010}. \nh{Despite its mechanical softness, the present description remains within the small-strain regime, which is required to justify the reduction to a single buckling coordinate.}

To estimate the effective buckling potential $V_{\text{eff}}(b)$, we performed ab initio calculations in the framework of the density functional theory using the Vienna Ab initio Simulation Package (VASP) \cite{vasp}. The generalized gradient approximation of Perdew, Burke and Ernzerhof \cite{perdew1996generalized} was employed with a cut-off energy of 400~eV. For the unstrained configuration, we relaxed the bdc molecule to its computational ground state, which is $\approx 4.5\%$ larger than experimentally observed in MOF-5~\cite{Lock2010}. \nh{For the analysis, we introduce a coordinate system $(x,y,z)$ attached to an individual linker.} In all calculations, the linker axis was aligned along the $x$-direction, while the benzene ring was constrained to the $xy$-plane. The displacement due to buckling was applied along the $z$-direction, perpendicular to the molecular plane. We then decreased the molecular length successively and calculated the total energy for a sine-shaped displacement with varying amplitude. The energy against the buckling mode coordinate is shown in Fig.~\ref{fig:potential}(c) for an external strain up to 4\%. It can clearly be seen that the total energy (effective potential) takes the form of a double-well potential, which can be approximated by $V_{\text{eff}}(b) = -\mu\, b^2 + \nu/2\, b^4$. For sufficiently high strain $\geq 2\%$, we find $\mu > 0$ and a corresponding non-zero minimum in the buckling coordinate. 

\begin{figure}[t]
    \centering
    \includegraphics[width=0.48\textwidth]{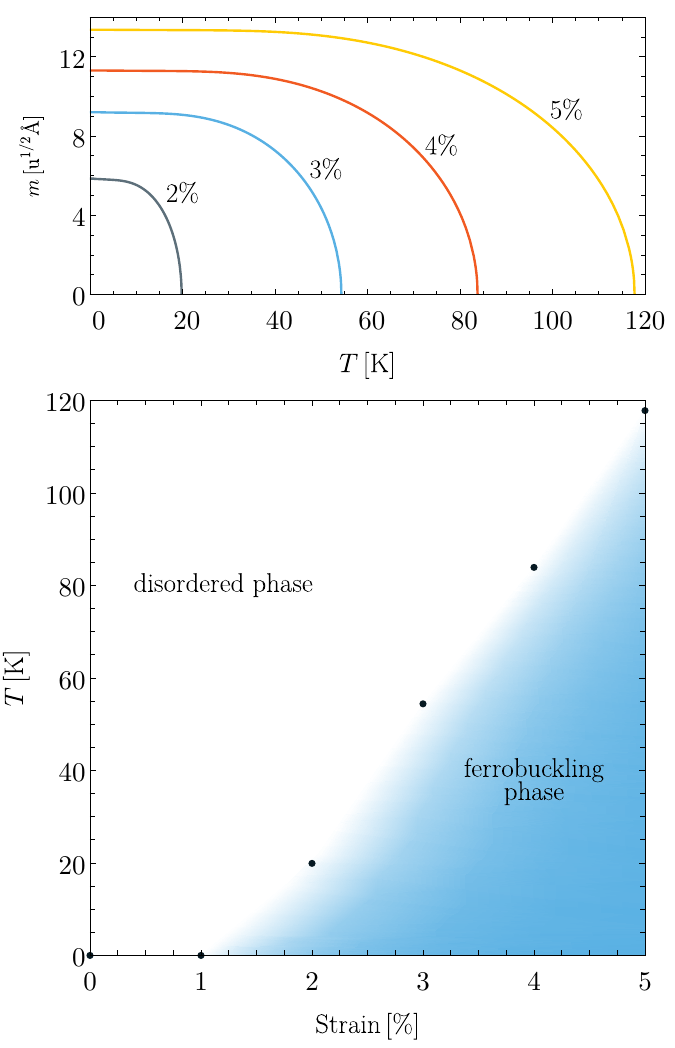}
    \caption{Mean-field order parameter $m$ as function of temperature for different values of applied strain. Phase diagram in the temperature-strain plane, showing a transition between a ferrobuckling phase and a disordered phase.}
    \label{Fig:PhaseDiagram}
\end{figure}

To quantify the coupling between the linkers, we further computed the Born effective charge \eqref{3EffCharge} from the response of the dipole moment to a sinusoidal displacement of the bdc linker. Choosing the soft buckling mode $v^{(1)}$ as a sine and accounting for the atomic masses, we obtain a mode dipole vector of $d_z = 8.60869\times 10^{-2} e/\sqrt{\text{u}}$, see Eq.~\eqref{3DipoleApprox}. This quantity is independent of strain, in contrast to the parameter $\mu$ and $\nu$ that determine the effective potential.

To keep the present work concise, \nh{we restrict the analysis to decoupled (100)-planes of MOF-5. This constitutes a controlled simplification that allows us to isolate the essential physics of collective buckling within a minimal setting. In a full three-dimensional description, inter-plane couplings are generally present and may modify the resulting ordering patterns.} We obtain a mean-field coupling parameter of
\begin{equation}
    \Tilde{J} = 58.09\, \frac{\mu \text{eV}}{\text{u}\, \text{\AA}^2},
\end{equation}
\nh{where we used the vacuum permittivity $\epsilon = \epsilon_0$ for simplicity.}

In Fig.~\ref{Fig:PhaseDiagram} we show the order parameter as obtained by a numerical solution of the self-consistency equation as a function of temperature and the phase diagram. The model parameters and the critical temperature are summarized in Tab.~\ref{Tab1}. The mean-field analysis predicts a continuous transition into a ferrobuckled phase and the critical temperature increasing approximately linearly with applied strain. For $T \to 0$ the order parameter saturates at $b_0 = \sqrt{\mu/\nu}$, the minimum of the double-well potential. \nh{These results suggest, within the present model framework, that collective buckling could occur at experimentally accessible temperatures under moderate strain.}

\begin{table}[ht]
    \centering
    \renewcommand{\arraystretch}{1.5}
    \begin{tabular}{|c|c|c|c|}
        \hline
        \hspace{0.2cm} \text{Strain [\%]} \hspace{0.2cm} & \hspace{0.1cm}
        $\mu\,[\text{meV}/(\text{u}\,\text{\AA}^2)]$ \hspace{0.1cm} & \hspace{0.1cm}
        $\nu\,[\mu\text{eV}/(\text{u}^2\,\text{\AA}^4)]$ \hspace{0.1cm} & \hspace{0.2cm}
        $T_C\,[\text{K}]$ \hspace{0.2cm} \rule{0pt}{3ex} 
        \\
        \hline\hline
        0 & $-0.94$ & 10.30 & 0 \\
        \hline
        1 & $-0.30$ & 11.66 & 0 \\
        \hline
        2 & \phantom{$-$}0.42    & 13.22 & 19.80 \\
        \hline
        3 & \phantom{$-$}1.24    & 15.00 & 54.34 \\
        \hline
        4 & \phantom{$-$}2.13    & 16.91 & 83.80 \\
        \hline
        5 & \phantom{$-$}2.70    & 15.29 & 117.75 \\
        \hline
    \end{tabular}
    \caption{Model parameters and critical temperature as functions of applied strain.}
    \label{Tab1}
\end{table}

\section{Classical to Quantum Cross-Over}
\label{secVI}

The crossover from classical to quantum buckling is governed by the local quantization energy of a single linker. This scale is obtained from the harmonic approximation around one of the buckling minima. For the effective potential \eqref{2EffPot} it is
\begin{equation}
    V_{\text{eff}}(b) = -\frac{\mu^2}{2 \nu} + \frac{\omega_1^2}{2} (b-b_0)^2 + \mathcal{O} \left( (b-b_0)^4 \right),
\end{equation}
where the two degenerate minima and the local well frequency are given by
\begin{equation}
    b_0 = \pm \sqrt{\mu/\nu},
    \qquad
    \omega_1 = \sqrt{V''_{\text{eff}}(b_0)} = 2\sqrt{\mu}.
\end{equation}
Once the thermal energy becomes comparable to or smaller than this scale
\begin{equation}
    k_\mathrm{B} T \lesssim \hbar \omega_1,
\end{equation}
each linker effectively behaves as a two-level system described by the two lowest eigenstates of the double-well potential. The crossover temperature is therefore
\begin{equation}
    T_1 = \hbar \omega_1/ k_\mathrm{B}.
\end{equation}

The total energy of a single linker is
\begin{equation}
    E = \frac{1}{2} \dot{b}^2 + V_{\text{eff}}(b)
\end{equation}
and for unit effective mass the canonical momentum of the buckling coordinate is $p_b = \dot{b}$. Canonical quantization $p_b \to -i\hbar \partial/\partial b$ yields the Schrödinger equation
\begin{equation}    \label{6Schrodinger}
    \left[ -\frac{\hbar^2}{2} \frac{\partial^2}{\partial b^2} + V_{\text{eff}}(b) \right] \psi(b) = E \psi(b)
\end{equation}
with a buckling wave function $\psi(b)$.

\begin{figure}[t]
    \centering    \includegraphics[width=0.45\textwidth]{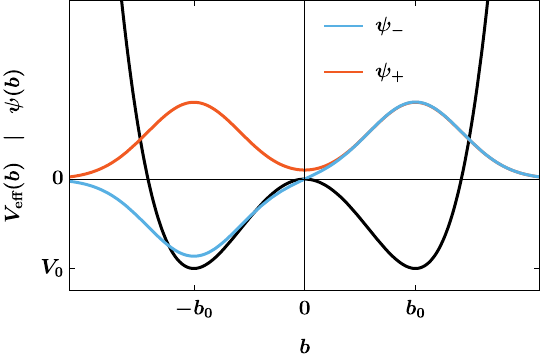}
    \caption{Effective potential and wave functions of the ground state $\psi_-$ and the first excited state $\psi_+$.}
    \label{Fig:BucklingWaveFunction}
\end{figure}

In the limit of a large potential barrier, the two lowest eigenstates are strongly localized near $b_0$. Each well can be approximated by a harmonic oscillator with ground state energy $E_0 = V(\pm b_0) + \hbar \omega_1/2 = -\mu^2/(2\nu) + \hbar \sqrt{\mu}$. Lowering the potential barrier enables tunneling between the two wells and lifts the degeneracy
\begin{equation}
    E_{\pm} = E_0 \pm t,
\end{equation}
where $t > 0$ is the tunneling amplitude. Due to the symmetry of the potential, the wave functions are either even or odd in $b$. Sturm-Liouville theory guarantees that the ground state is even, while the first excited state is odd \cite{TeschlBook}. These states are well approximated by symmetric and antisymmetric combinations of the localized harmonic oscillator ground states. An effective Hamiltonian for this two-level system is
\begin{equation}
    h = \begin{bmatrix}
        E_0 & -t
        \\
        -t & E_0
    \end{bmatrix}
\end{equation}
with eigenstates $\psi_\pm = 1/\sqrt{2} ( 1, \pm 1)$ and energy splitting $2t$. Fig. \ref{Fig:BucklingWaveFunction} shows the double-well potential \eqref{2EffPot} and the wavefunctions of the ground state $\psi_+$ and the first excited state $\psi_-$. 

The tunnel amplitude $t$ can be estimated semiclassically within the WKB approximation \cite{LanLifshitz}
\begin{equation}
\begin{split}
    t &\approx \frac{\omega_2 \hbar}{2 \pi} \exp \left( -\frac{1}{\hbar} \int\limits_{-b_1}^{b_1} db \abs{p(b)} \right)
    \\
    &= \frac{\omega_2 \hbar}{2 \pi} \exp \left( \pi \frac{E_0}{\hbar \omega_2} + \frac{\pi}{\sqrt{2}} \right),
\end{split}
\end{equation}
where $p(b)$ is the momentum in the classically forbidden region, $\pm b_1$ are the turning points and $\omega_2 = \sqrt{2 \mu}$ the barrier frequency. To reproduce the result in Ref.~\onlinecite{Savelev2006} we applied a harmonic approximation of $V_{\text{eff}}(b)$ around the local maximum. Once the thermal energy becomes comparable to or smaller than this scale, quantum tunneling dominates over thermal hopping. This defines a second temperature scale $T_2 = t/k_\mathrm{B}$ with $T_2 \ll T_1$. 

Below this temperature, $t$ becomes relevant and we write the local Hamiltonian in the two-level subspace of $\psi_\pm$ as $-t\, \sigma^x$. It remains to couple the linkers according to the results of Sec. \ref{secIII}. Substituting $b_i \to b_0 \sigma_i^z$ yields the transverse field Ising model
\begin{equation}
    H = - \frac{b_0^2}{2} \sum_{\langle i,j\rangle} J_{ij} \sigma_i^z \sigma_j^z + t \sum_{i} \sigma_i^x.
\end{equation}
This model has been discussed in the context of collective buckling in Ref.~\onlinecite{Geilhufe2021}. The competition of the coupling $J_{ij}$, favoring an ordered state, and the transverse field $t$ leads to a quantum phase transition into the parabuckling state.

We estimated $t$ by the WKB approximation for MOF-5 and found \nh{negligibly} small values. This is in accordance with a numerical solution of the Schrödinger equation \eqref{6Schrodinger}, which shows no energy splitting between the two lowest eigenstates. We conclude that a quantum crossover is absent for the present example.

\section{Conclusion}
\label{secVII}

We have developed a framework to describe collective buckling in MOFs. Starting from the microscopic structure of a single organic linker, we defined a buckling coordinate governed by an effective double-well potential. The coupling between linkers was derived from a dipole-dipole approximation, resulting in \nh{an effective lattice Hamiltonian \eqref{1Ham}}.

We introduced a ferrobuckling order parameter and analyzed the phase transition within a mean-field approximation. Applying our theory to MOF-5 \nh{as an illustrative example}, we extracted the model parameters from DFT and obtained the critical temperature as a function of applied strain for a decoupled (100) plane. At low temperatures, quantum effects may drive a crossover to a transverse-field Ising model, giving rise to a parabuckling phase.

\nh{The present theory constitutes a coarse-grained description that focuses on a single symmetry-breaking deformation mode of the framework. Other low-energy distortions, such as collective displacements of the metallic coordination centers that are geometrically coupled to angular distortions of the network, are not treated explicitly but are expected to accompany linker buckling in realistic materials. In this sense, the buckling coordinate provides a minimal and complementary description of collective framework deformations.}

\nh{While the collective phases were analyzed within a mean-field approximation, the long-range nature of the dipolar interaction suggests that the qualitative structure of the phase diagram is robust. Fluctuations beyond mean field are expected to renormalize quantitative details such as transition temperatures without altering the nature of the ordered phases \cite{AltlandSimons}.}

\nh{Our work provides a quantitative starting point for describing macroscopic buckling behaviour in terms of microscopic degrees of freedom.} \nh{This approach can be extended} beyond decoupled (100) planes to the full three-dimensional structure of MOF-5, allowing for more general order parameters and potentially richer collective phases. \nh{Experimental identification of collective buckling phases in MOFs, as well as their coupling to electronic degrees of freedom, remains an open challenge for future work.}

\begin{acknowledgments}
All authors acknowledge support from the Areas of Advance Nano and Materials Science at Chalmers University of Technology. NH and RMG acknowledge support from the Swedish Research Council (VR starting Grant No. 2022-03350), the Olle Engkvist Foundation (Grant No. 229-0443), the Royal Physiographic Society in Lund (Horisont), and the Knut and Alice Wallenberg Foundation (Grant No. 2023.0087). The computations were enabled by resources provided by the National Academic Infrastructure for Supercomputing in Sweden (NAISS) at C3SE, NSC, and PDC, partially funded by the Swedish Research Council through grant agreement no. 2022-06725.
\end{acknowledgments}

\bibliography{lit}

\appendix

\section{Free Energy Expansion in Mean-Field Approximation}
\label{A1}

The free energy of the mean-field Hamiltonian \eqref{4HamMF} is given by
\begin{equation}
    F = - k_{\mathrm{B}} T \ln\, Z = \frac{N \Tilde{J}}{2} m^2 - k_{\mathrm{B}} T N \ln\, Z_{\text{loc}},
\end{equation}
where the local partition function is
\begin{equation}
    Z_{\text{loc}} = \int db\, e^{-\beta\, h(b)}.
\end{equation}
Instead of the free energy, we consider in the following the free energy density $f = F/N$, which is an intensive quantity. Adopting the notation of the main text \eqref{4SinglePartPartFunc}, we write \nh{the local partition function} as
\begin{equation}
    Z_{\text{loc}} = Z_0 \left\langle e^{\beta \Tilde{J} m b} \right\rangle_0.
\end{equation}
Close to the phase transition the order parameter $m$ is small and we can perform a cumulant expansion \cite{Kubo1962}
\begin{equation}    \label{ACumulantExpansion}
    \ln \left\langle e^{\beta \Tilde{J} m b} \right\rangle_0 = \sum_{n=1}^\infty \frac{(\beta \Tilde{J} m)^n}{n!} \kappa_n = \sum_{n=1}^\infty \frac{(\beta \Tilde{J} m)^{2n}}{(2n)!} \kappa_{2n}
\end{equation}
with the cumulants $\kappa_n$. Since $V_{\text{eff}}(b)$ is even, odd terms vanish. 

The partition function $Z_0$ for a single linker has an analytical solution ($a = \beta \nu/ 2 > 0$, $c = \beta \mu \in \mathbb{R}$)
\begin{equation}
    Z_0 = \int db\, e^{-a b^4 + c b^2} = \sqrt{\pi}\, (2a)^{-1/4} e^{c^2/(8a)} D_{-1/2} \left( -\frac{c}{\sqrt{2a}} \right)
\end{equation}
in terms of parabolic cylinder functions $D_n(z)$, which are a generalization of the Hermite functions to non-integer order \cite{NIST}. The moments can be obtained from the derivatives
\begin{equation}
    \left\langle b^{2n} \right\rangle_0 = \frac{1}{Z_0} \frac{\partial^n}{\partial c^n}  Z_0
\end{equation}
\nh{and the cumulants from the standard generating function}
\begin{equation}
    \nh{\kappa_n = \frac{\partial^n}{\partial t^n} \ln \left\langle e^{t\, b} \right\rangle\Bigg|_{t=0}.}
\end{equation}
To obtain a recursion formula for the moments, we compute the derivative
\begin{equation}
\begin{split}
    &\frac{d}{db}\left( b^{2n+1} e^{-a b^4 + c b^2} \right) = 
    \\
    &\left[ (2n+1)b^{2n} -4a\, b^{2n+4} + 2c\, b^{2n+2} \right] e^{-a b^4 + c b^2}
\end{split}
\end{equation}
and integrate over the real numbers. This yields
\begin{equation}
    \left\langle b^{2n+4} \right\rangle_0 = \frac{c}{2a} \left\langle b^{2n+2} \right\rangle_0 + \frac{2n+1}{4a} \left\langle b^{2n} \right\rangle_0.
\end{equation}
Therefore it suffices to compute the derivative for $n = 1$
\begin{equation}
\begin{split}
    \langle b^2 \rangle_0 =& \frac{c}{2a} + \frac{1}{\sqrt{2a}} R\left( -\frac{c}{\sqrt{2a}} \right)
    \\
    =& \frac{\mu}{\nu} + \frac{1}{\sqrt{\beta \nu}} R \left(- \mu \sqrt{\frac{\beta }{\nu}} \right),
    \\
    R(z) =& \frac{D_{1/2}(z)}{D_{-1/2}(z)}.
\end{split}
\end{equation}
We also explicitly state the fourth moment
\begin{equation}
\begin{split}
    \langle b^4 \rangle_0 =& \frac{1}{4a} + \frac{c^2}{4a^2} + \frac{c}{(2a)^{3/2}} R \left( -\frac{c}{\sqrt{2a}} \right)
    \\
    =& \frac{1}{2 \beta \nu} + \frac{\mu^2}{\nu^2} + \frac{\mu}{\beta^{1/2} \nu^{3/2}} R\left( -\mu \sqrt{\frac{\beta}{\nu}} \right).
\end{split}
\end{equation}
At low temperatures we can use the asymptotics of the parabolic cylinder functions for $z \to \infty$ to find $R(z) = 1/(2z) + \mathcal{O}(z^{-3})$. This coincides with a saddle-point approximation around both minima of the double-well.

The cumulant expansion \eqref{ACumulantExpansion} up to $\mathcal{O}(m^4)$ yields
\begin{equation}
    f = -T \ln Z_0 + \frac{a_2(T)}{2} m^2 + \frac{a_4(T)}{4} m^4 + \mathcal{O}(m^6)
\end{equation}
with the temperature-dependent Landau coefficients
\begin{equation}
    a_2(T) = \Tilde{J} \left( 1 - \beta \Tilde{J} \kappa_2 \right),
    \qquad
    a_4(T) = -\frac{\beta^3 \Tilde{J}^4}{6} \kappa_4.
\end{equation}
The first two non-vanishing cumulants are given by
\begin{equation}
    \kappa_2 = \langle b^2 \rangle_0,
    \qquad
    \kappa_4 = \langle b^4 \rangle_0 - 3 \langle b^2 \rangle_0^2.
\end{equation}
A change of sign in $a_2(T)$, i.e. its root, as a function of temperature determines the phase transition. We arrive at
\begin{equation}
    k_{\mathrm{B}} T_C = \Tilde{J} \langle b^2 \rangle_0,
\end{equation}
which is to be understood as a fixed-point equation as the right-hand side depends on temperature as well. By minimizing the free energy, we obtain an expression for the order parameter
\begin{equation}
    m(T) = \sqrt{-\frac{a_2(T)}{a_4(T)}} = \sqrt{\frac{6}{\beta^3 \Tilde{J}^3} \frac{1 - \beta \Tilde{J} \kappa_2}{\kappa_4}},
\end{equation}
which is valid only close to the phase transition. A first order expansion around $T_C$ reveals the characteristic mean-field behaviour of the order parameter
\begin{equation}
    m(T) \sim \sqrt{T_C - T},
\end{equation}
i.e. the critical exponent is $1/2$. The condition $a_4 > 0$ guarantees that the expansion is stable and that the phase transition is continuous.

\end{document}